# The nested materiality of environmental monitoring

*Elena Parmiggiani and Eric Monteiro*

Department of Computer and Information Science
Norwegian University of Science and Technology
Trondheim, Norway

{parmiggi;ericm}@idi.ntnu.no

**Abstract**. Present knowledge about the marine ecosystem on the Norwegian Continental Shelf towards the Arctic is sparse. These areas are vast, remote and subject to harsh weather conditions. We report from a three-year case study of an ongoing effort for real-time, subsea environmental monitoring by an oil and gas operator. The 'facts' about the subsea environment are anything but neutral; they are intrinsically caught up with the material means by which they are known. The marine ecosystem is monitored through a network of sensors, communication links, visualisation and analysis tools. Our concept of nested materiality draws heavily on perspectives in sociomateriality but highlights (i) the distributed and interconnected infrastructure of the material means (as opposed to artefact-centric) and (ii) in-the-making (as opposed to black-boxed) technology.

**Key words**: sociomateriality, performativity, environmental monitoring, infrastructure.

## 1. Introduction

With the 'easy' oil already found, oil and gas operations in high-cost, climatically challenging, offshore locations like the Norwegian Continental Shelf (NCS) are knowledge-intensive. Oil reservoirs on the NCS reside 3 – 5 kilometres below the seabed and are known largely through echogram reflections from hydrophones (i.e., seismics). More than 50% of hydrocarbons on the NCS are produced by unmanned subsea installations placed on the seabed. Operational decisions rely on sensor data streams of pressure, temperatures, choke positions, sand detection, and flow volumes, fed by fibre-optic networks and visualised in onshore control rooms. The necessary 'facts' for safe and efficient operations, then, are anything but neutral: they are intrinsically caught up with their material means (sensors, networks, simulations, visualisation) by which they become known (Latour 1999; Almklov 2008).

This general insight is particularly evident when studying knowledge about a quite new (to the oil and gas sector) domain viz. subsea environmental monitoring. Commercial interests in oil and natural gas are pushing north towards the Arctic, into presently banned areas. Extreme weather conditions and a precarious environment make oil operations highly controversial. A political lifting of the ban hinges on establishing a robust 'knowledge base' (NME 2011). Given the sparse, existing knowledge of the marine ecosystem in and close to the Arctic (Blanchard et al. 2014), our three-year case study reports on NorthOil's (a pseudonym) ongoing efforts to establish a capacity for real-time subsea environmental monitoring. We ask: *how are facts about the subsea environment produced?*

We draw on insights from sociomateriality about the constitutive entanglement of technology, work, and knowledge (Orlikowski and Scott 2008; cf. special issue of MIS Quarterly by Cecez-Kecmanovic et al. 2014; and the Scandinavian Journal of IS by Bratteteig and Verne 2012). We adopt a performative rather than representational approach (Pickering 1992). The material circumstances of how 'facts' are produced are crucial yet often black-boxed (Orlikowski and Scott 2014; Østerlie et al. 2012; Pollock 2012). The capacity for environmental monitoring we study is in-the-making thus provides an occasion to open the black box of how key choices are made: *what* aspects to select (e.g., fish, eggs, water, corals), *how* to measure (e.g., echo sounders, pictures, video), *when* to sample (minutes, hours, days), how to represent and *visualise* (e.g., aggregates, graphs, pictures).

We contribute to the modest but growing stock of studies in sociomateriality demonstrating the performativity (Kallinikos and Tempini 2014; MacKenzie and Millo 2003; Østerlie et al. 2012; Pollock 2012) rather than merely proclaiming it (cf. Cecez-Kecmanovic et al. 2014; Jones 2014). More specifically, our notion of nested materiality highlights performativity as *distributed*, *interconnected,* and *interacting* rather than through any singular artefact or sensor alone; nested materiality demonstrates the general principle of performativity found in sociomateriality within the *infrastructure* for environmental monitoring (Jensen and Winthereik 2013; Monteiro et al. 2013).

## 2. Perspectives on sociomateriality

### 2.1 Background and precursors to sociomateriality

The discourse on how to conceptualise technology runs long in information systems research. As a counter-reaction to overly deterministic accounts, the significant discretion for users to appropriate information systems was established decades ago through empirical studies (Barley 1986; Gasser 1986; Kling 1986) as well as theoretical concepts (e.g. the 'situated' nature of action proposed by Suchman (1987), the presence of 'workarounds' by Gasser (1986) and leaning on Giddens' structuration theory as proposed by Orlikowski and Robey (1991) and Walsham (1993)).

In their historical recapitulation, Orlikowski and Scott (2008) describe three, broad categories of approaches: (i) discrete entities (with uni-directional causal effects of technology), (ii) mutually dependent ensembles (with bi-directional relationship), before outlining (iii) sociomaterial assemblages. While (i) come with overly deterministic connotations, Orlikowski and Scott (ibid.) acknowledge the rich source of insights provided by (ii). Especially influential are practice-based perspectives (cf. also Jones 2014).

Practice-based research demonstrates the significant malleability of the use of technology by, in a given context, identifying both intentional and unintentional changes resulting from local appropriation, workarounds, and situated innovation that go into users' enactment of technology (Gherardi 2006; Suchman 1987). The use of information systems, then, is malleable because 'every encounter with technology is temporally and contextually provisional, and thus there is, in every use, always the possibility of a different structure being enacted' (Orlikowski 2000, p. 412). A user, accordingly, has substantial freedom to enact her practices with technology in different ways. This malleability in the use of technology enables us to resolve paradoxical or contradictory empirical findings regarding different outcomes of the same technology: seemingly contradictory outcomes are simply a result of contextual differences (Barley 1986; Robey and Boudreau 1999).

The decisive distinction between the former two approaches and sociomateriality, Orlikowski and Scott (2008, p. 455 emphasis added) point out is that:

> "[sociomateriality] is a move away from focusing on how technologies influence humans, to examining how materiality is intrinsic to everyday activities and identities (...) material means are not so much tools to be used to accomplish some tasks, but *they are constitutive of both activities and identities*".

## 2.2 The performative turn

As noted further by Orlikowski and Scott (2008, p. 460), "[a] central idea entailed in sociomateriality is the notion of performativity". Performativity is an operationalization of the constitutive entanglement of the material and the social (see also Cecez-Kecmanovic et al. 2014; Jones 2014) But what does this 'performativity' entail?

In a widely cited study of the financial option market, Mackenzie and Millo (2003) explicitly set out to demonstrate the performativity of certain formula (the so-called Black-Scholes model) by showing how its initially descriptive role gradually got replaced by an enacting role when the formula was inscribed in (trading) robots and professional routines. As MacKenzie and Millo (ibid., p. 107, cited in Orlikowski and Scott (2008, p. 461)) note: "Option pricing theory (…) succeeded empirically not because it discovered pre-existing price patterns but because markets changed in ways that made its assumptions more accurate and because the theory was used in arbitrage".

However, despite repeated calls to eliminate the dichotomy between the social and the material (Orlikowski and Scott 2008), "the social almost always seems to take precedence, the material merely affording some social/human intention" (Cecez-Kecmanovic et al. 2014, p. 861). Jones (2014, p. 922) too notes that despite claims of the opposite, actual demonstrations "seem [to be] only selectively recognized in the extant literature". Hence the slogan "materials matter" (cf. Barad 2003) is sometimes exactly that, a slogan. The detailing of *how*, not *that*, materials matter (viz. their performativity) remains under-specified.

However, notable exceptions exist (e.g., Jones 2014; Kallinikos and Tempini 2014; Orlikowski and Scott 2014; Østerlie et al. 2012; Pollock 2012). Based on a case study of TripAdvisor, Orlikowski and Scott (2014) show that the nature of a service at any time and place "reflects the materiality involved in its constitution in practice (e.g., equipment, medial, channels, bodies, buildings, spaces, etc.)" (p. 4 in preprint) Similar to the Black-Scholes model, the algorithms at the hearth of TripAdvisor constantly configure – or materialize – the services by conveying choices about what to exclude or not and what should be left explicit or implicit. On the one hand, by materializing the service, algorithms powerfully shape the practices of user crowds. On the other hand, the user crowd also plays a vital role in producing the content configured by the algorithms.

## *2.3 Towards nested materiality*

Particularly relevant to us are the studies of sociomaterial knowing i.e. demonstrations of performativity specifically targeting the production of 'facts'.

Kallinikos and Tempini (2014) analyse how medical knowledge can be created and organized into new models where social media platforms play a performative role (cf. Treem and Leonardi 2012). Social media platforms are complex technological arrangements where social relations are built and shaped by the computational operations embedded into the systems. The authors draw on the case of a social network for medical research and show how. In this process new 'facts' (viz. new correlations between the life paths of patients thus new knowledge for doctors) are materialized through data manipulations where patients are dynamically linked with other patients via the intermediation of a carefully structured architectural underpinning of the particular social media platform. Essential to medical knowledge production is the specific amalgamation of data architecture and computational capabilities with the user interactions, which mutually constitute each other.

Pollock (2012) studies how industry analysts like Gartner produces 'facts' about the market situation of technology vendors in different business domains. Ranking devices, Pollock demonstrates, are performative and ultimately change market domains through the rankings. The ranking device introduces changes to a market domain so that it fits the ranking produced, quite similar to the Black-Scholes model.

In a study empirically close to ours, Østerlie et al.'s (2012) study how 'facts' about non/presence of sand in the oil and gas stream are produced. Presence of sand may severely damage processing equipment. Østerlie and colleagues' study the monitoring work of petroleum engineers through software applications linked to the sand detection sensors installed along the well path. As the well flow is digitalized and becomes a data stream visualized on monitors, 'facts' about sand result from practices, material arrangements for inspecting phenomena, and the physical characteristics of sand. The authors also show how it is crucial in the engineers' daily work to often proceed backwards and unpack this construction process to detect malfunctions and solve anomalies.

Our study clearly shares deep affinities with these studies of 'fact' production. The notion of nested materiality draws on these studies of sociomaterial knowing but is a vehicle to highlight (i) the *distributed* and *interconnected* performativity of the subsea environmental monitoring infrastructure along a punctuated network of digital devices (as opposed to more artefact-centric focus on Gartner's Magic quadrant device or the sand detector sensor) and (ii) the design choices of environmental monitoring in-the-making (thus by *interacting*, as opposed to technology that is black-boxed for the users).

## 3. Case setting and method

### *3.1 Case setting*

Oil and gas activity on the NCS has expanded dramatically since its inception more than forty years ago. From a modest start, relying heavily on foreign (notably US) expertise, it is today a dominating industry in Norway employing (directly and indirectly) 10% of the workforce, accounting for 30% of GNP and 50% of net exports. Throughout this period, there have been tensions and conflicts with the traditional fishing industry as well as broader environmental concerns. Presently, the inherent conflicts are actualised by the ongoing controversy over whether to allow oil and gas operations in new areas in the Arctic North. These areas, the oil industry argues, are particularly interesting geologically but are also where the most commercially important fishing takes place. Moreover, the areas have stunning scenery, seeing numerous tourists and have rich environmental ecosystems.

Our case unfolds against this backdrop of heated debate. Our empirical material reports from NorthOil's (a pseudonym for a significant, internationally operating oil and gas operator) ongoing efforts to better position themselves for environmental demands and requirements – still not defined – expected to pertain to possible oil and gas activities in environmentally contested areas in the Arctic North. Given the profound lack of robust knowledge about the status of the environment, NorthOil with collaborators (including research institutions) have embarked on projects including the one we report on that aim at contributing towards a more robust baseline measurement of (selected aspects of) the environment.

## *3.2 Approach and access to case*

We document a case study of the establishment in NorthOil of a knowledge base about the marine ecosystems. This effort is facilitated by the introduction of technologies and methodologies for subsea real-time data collection, transfer, and visualization organized into an infrastructure for environmental monitoring.

In offshore oil and gas operations human physical access to the subsea site is impossible, so interaction is always mediated by digital technologies, varied in content, and distributed in space (not only from subsea to shore but also across different nations) and time (obtaining a baseline of environmental behaviour might require decades). Importantly, the installation of digital technologies is fairly recent and ongoing (Henderson et al. 2013) – a feature that allows us to inquire into the design choices as they are made.

The oil and gas business is traditionally secretive, so access was dependent on a number of pragmatic conditions. Facilitated by a member of our research group who also holds a position as project manager in NorthOil, we were introduced to the company's Norwegian research centre, where the first author could follow an ongoing three-year project (December 2011 - December 2014) to set up an infrastructure for real-time environmental monitoring in collaboration with a number of vendors and project partners. The second author has a history of collaboration with NorthOil.

## *3.3 Data collection*

The collection of empirical data was conducted through an ethnographic method (Ribes 2014). We were granted access to NorthOil research centre in April 2012, where the first author spent on average two to three working days a week until April 2014. Main sources of data generation were: participant observations, interviews, documents, and corporate information systems. In Table 1 is a summary of the data types, their amount, and the topics covered.

-- Table 1 about here --

The first author was given a badge and a desk in a shared office space, where she could follow the ongoing activities, join in meetings, workshops, and teleconferences with the industrial partners. In case circumstances did not allow for note taking, the relevant points were transcribed as soon as possible. The regular presence of the first author in the research field also made it possible to shift from being considered an outsider to one of NorthOil's employees. As acknowledged by Klein and Myers' principle of interaction between researcher and the subjects (1999), the continuous, informal contact let us obtain more and richer kinds of data.

Participant observations were crucial to identify the main informants for collecting semi-structured interviews, also from the partner companies. 33 interviews were conducted lasting on average 1 hour, mostly audio recorded and subsequently transcribed. In 8 cases we had no permission or chance to tape, so notes supplemented the lack of transcription.

We also had access to public and restricted documentation (including email threads, slides, minutes of meetings, reports). Together with interviews, documentation was fundamental to understand the technical characteristics and the setup of the subsea sensors and devices adopted for environmental monitoring. In addition, through the company's intranet, we could look at and (sometimes partially) make use of the same information systems used by the employees, e.g., in-use or test modelling software and web portals to track real-time environmental data. These systems were seen or tested by either the first or both authors during meetings or as access was granted to them. Documentation describing these systems was also retrieved and analysed. Interviews and participant observations constituted a backdrop for the identification and interpretation of internal documentation and systems.

The second author occasionally participated in meetings, interviews, and also had access to restricted documentation.

## 3.4 Data analysis

Our object of study is NorthOil's ongoing efforts to implement a new infrastructure composed of communication architecture and methods to support real-time subsea environmental monitoring. These efforts involve developing, testing, and integrating a large number of sensors, tools, methodologies, and organisational routines. However, the organisational roles of the end users and the decision gates are only partly clear to NorthOil's employees taking part in the initiative. Envisioned work practices include supporting environmental advisors in deciding when the spawning season in contested areas should be halted. This relies on the ability to capture, format, analyse, and present the concentration of environmental resources in the area (e.g., spawning fish and coral reefs). In addition, online risk analysis for static environmental resources like corals should be provided to drilling engineers when a new oil well is drilled on the NCS. If the discharges are transported too close to corals by the stream, any drilling activity must be stopped immediately. The risk analysis capacities are mostly provided by general-purpose semi-automatic modelling software to simulate the dispersion of particles in the water or to analyse pictures of the marine resources like coral structures. Therefore, our unit of analysis is the early stages of infrastructure design and development that precede adoption by the oil and gas professionals in their daily tasks.

Our data analysis is guided by Klein and Myers' (1999) principles for interpretive research. Data analysis was iterative and interleaving inductive and deductive steps. We began inductively by open coding our field notes, interview transcripts, and documentation in parallel with the data collection. The temporal overlap with data collection was particularly significant with reference to our results. For NorthOil's experts the relationship between marine environmental knowledge and the collection of real-time data sets was highly entangled. Consider the example of fish migrations tracking in the NCS. Traditional fish migrations monitoring is conducted from boats. Using the same devices, NorthOil decided to conduct online monitoring from the seafloor through immobile monitoring station that would grant the collection of consistent datasets. No method, however, was known to NorthOil and its advisors to directly compare the datasets collected with the new method with the historical datasets acquired with the old top-down approach. The underlying reason to the mismatch is that the fish's swim bladder reflects the sensor beams differently when hit from different angles (in this case, from below rather than from above). The causal relationship between results, instruments, and the data translation process (the fact) was thus being explored by NorthOil's project participants themselves. We saw this indeterminacy as an opportunity: it allowed us to see the centrality of the problem of performativity of fact production.

Next, a striking observation in our fieldwork was how the facts (e.g. risk for coral reef) emerging from *distributed* and *interconnected* arrangements of natural elements (e.g., coral structures), sensing devices (e.g., a subsea camera), and risk analysis software (e.g., a particle spreading model). Rather than singular sensors like that above to capture fish bladders, environmental facts emerged through an infrastructure. Our resulting interpretative template (see Table 2) spans three moments along the distributed, interconnected infrastructure from the seafloor ('nature') via operations tied to production and maintenance ('operational site') and, finally, visualisations of environmental risk for users.

-- Table 2 about here --

# 4. Findings

## *4.1 Performing nature*

*Spotting the coral reefs*. The seas off the Lofoten islands in the Arctic areas of Norway are currently prohibited to oil and gas operations due to the density and richness of their natural ecosystems. They are in particular inhabited by many cold-water coral reefs and host dense migration and spawning of commercially relevant fish species like cod and herring. In the mid-2000s NorthOil deployed a subsea lander (a semi-conic metallic structure, Figure 1) equipped with a sensor network on the seafloor offshore Arctic Norway. The lander was conveniently placed close to a coral structure inhabited by various marine species. In 2013, the lander was connected to an onshore data centre through a fibre-optic cable and the datasets collected by the sensors can be freely visualized and downloaded on a publicly accessible web portal.

The goal of NorthOil was to gain significant background knowledge about the environmental baseline in the area to show authorities the ability to operate without harming the marine environment, had they opened the area to oil and gas extraction. The Arctic lander is thus a key node for NorthOil to discover a non-operational, unknown area. Moreover, the availability of real-time data provides NorthOil environmental experts with a different lens: the subsea environment feels now closer and visible online. The current lander hosts a few off-the-shelf devices, for example *echo* sounders (acoustic devices able to identify obstacles, e.g. fish, in the water column); sensors to track *oceanographic* parameters (pressure, temperature, salinity); and a *camera* taking pictures every 30 minutes. The position of the lander and the orientation of the sensors are not arbitrary but the result of a long process of testing different approaches. The final position was such that the features of the coral structure could fit well in the camera lens.

This new perspective triggered discussions about the best sensor configuration to capture marine life around the coral reef. Initially installed to monitor the health of one coral structure, the subsea lander proved a valuable tool to also track the fish traffic above and around the coral. In this respect, the role of the echo sounders emerged as very important to differentiate the visible from the invisible. The two following snapshots exemplify that the fish are materialized on the users' desktop as the intersection between their own physical materiality, that of the sensors, and that of the modelling software plugged in to inject the missing datasets.

-- Figure 1 about here –

*The swim bladder.* Echo sounders send acoustic signals at given time intervals and measure the strength of the echo returned by the obstacle hit by their signal. Their performance depends on a number of parameters, for instance the density of the obstacle. Some fish (e.g., cod) have a dorsal swim bladder, an internal organ which is filled with air and allows the fish to control its buoyancy and to emit and receive sounds (Figure 2). Those fish reflect better the signals and are thus easier to detect. NorthOil and its partners soon realized that the direction of the acoustic signal with respect to the position of the swim bladder affects the interpretation of echo sounder data. Echo sounders are traditionally used by researchers and fishermen to detect fish by peeking downwards from ships or floating monitoring stations. However, NorthOil wanted to collect long-term data series from a static position, so boats were no longer an option. Placing the echo sounders on the seafloor means that the new measurements are instead taken upwards (Figure 3). Although expertise on how to compare and convert older and newer datasets is available to some vendors and research institutions, NorthOil's partners admitted it would require them several years to obtain either enough experience or the historical datasets to interpret the new measurements and relate them to existing knowledge bases.

*Adult fish vs. eggs and larvae.* The echo sounders installed on the Arctic lander are average commercial devices. Their hardware affects what they can detect because they cannot track objects smaller than their predefined wavelength (e.g., 2-4 cm). It is difficult if not impossible to track fish eggs (e.g., 1-2 mm) and larvae (e.g., 4-15 mm) that drift in the water column surrounding the coral structure (refer to Figure 3 for an illustration). However, given the high fish spawning activity in the area, NorthOil wanted to collect data series regarding the drifting of eggs and larvae. It was therefore decided to infer these data from general-purpose simulation models that work by abstracting larvae and eggs to particles following the water flow. Fish is instead more easily detectable (e.g., 5+ cm), but no models are available today to NorthOil's partners to describe fish movements. An environmental advisor points out:

> "*Each adult fish decides for itself!*"

One of her colleagues is frustrated by this situation, as the environmental experts are bouncing between having sensor data and only relying upon models to fill the gaps left by the missing data:

> "*Fish [are] detectable but we have no models; for larvae we have models but we cannot detect them!*"

-- Figure 2 and 3 about here –

## 4.2 Performing the site

The conditions that make subsea environmental monitoring possible proved site dependent. An environmental chemist explains that the materialization of a marine ecosystem emerges differently based on situated arrangements of the natural environment and the socio-political conditions of the area:

> *"[Environmental monitoring] needs to be based on the sort of political regime, basically the regulations in the country where you are operating. So maybe in some areas there might be a lot of seals or different species of fish, but if they're not a focus for the country, then you need to consider if this is still something that is relevant to monitor. And in other countries or areas it might be possible that a resource that is not really serving a key role in the ecosystem still has a high focus by the regulators, and that needs to be included. (...) [I]t's really a case-by-case decision"*

The Arctic lander was the only one NorthOil had in an area without operations, but other landers were installed in operational areas. For instance, one was allocated off the coast of Brazil, at a depth of more than 100 meters in an oil field which NorthOil had recently acquired. The sensors installed on the Brazilian lander were the same commodified, off-the-shelf devices used in the Arctic. Nevertheless, according to one environmental expert the construction of the lander and the configuration of the sensors had to "*dramatically change*". That was due not only to the presence of oil and gas activities but also to the characteristics of the Brazilian waters. The following empirical snapshots illustrate that environmental monitoring is performed not only within a specific entanglement of technologies and nature, but also within a broader socio-political background where interests and engagements play an important role.

*Different natural resources, different sensors.* The environment surrounding the Brazilian installation was quite different from the Arctic one. First of all, whereas coral reefs are very dense in the area around the Arctic lander, calcareous algae are the main inhabitants of the Brazilian oil field. Corals and algae are very different: whereas corals are animals, algae are plants and consequently need light to grow. The discharge of rock particles or the occasional leakages generated during oil and gas operations might increase the cloudiness (or turbidity) of the water. That is a critical problem for algae, as the particle cloud prevents the light from reaching the seafloor. As a result, NorthOil decided to also install a light sensor on the Brazilian lander. Another key difference between corals and algae is that the former can construct 35-meter-tall structures, whereas the latter lay on the seafloor in calcified structures the size of a golf ball (see Figure 4 and Figure 5). The camera used on the Arctic lander was installed on a 2-meter-high satellite crane unit cabled to the main lander. In the Brazilian case, the satellite crane was deemed unnecessary, so the camera was placed directly on the lander closer to the seafloor.

-- Figure 4 and 5 about here –

*Corrosion and marine snow.* The Brazilian waters are different from those in the Arctic because the former are warmer, more corrosive, and currents are generally much stronger. First, the corrosive effect came as a surprise to NorthOil, which was a rather new player in the region. The first steel lander deployed was in fact severely damaged and sensors failed due to corrosion and short circuits. A new lander made of titan was set out a few months later. This time sensors were better protected inside the lander. Second, the strong currents and the high temperature contribute to produce high density of so-called marine snow, mostly organic particles that are lifted from the seafloor and float by increasing the water turbidity. Marine snow does not reflect the sound waves well and therefore cannot be detected by the echo sounders installed on the Brazilian lander. Hence, workarounds had to be found to distinguish the water cloudiness caused by marine snow from the one generated by the more dangerous drilling discharges. A proposal was made to feed the camera pictures into software able to count the particles in the pictures and measure the amount of marine snow.

*Brazil vs. Arctic: Different countries require different focus.* The nesting of nature and technologies we have just exemplified is also the result of the socio-political conditions of the location where the nesting takes place. Besides, even if the environmental monitoring machinery had to be reconfigured to fit the Brazilian system, it was made possible by the situated experience NorthOil had acquired in the Arctic.

First, NorthOil had to comply with another very local element: the Brazilian authorities. While Norway is characterized by a tradition of well-established collaboration between authorities and oil and gas companies, there is a "*lack of cooperation between authorities and the industry world [in Brazil]*" (environmental advisor). The legal framework is also different, impacting on the speed of decisions related to drilling permits and approval of environmental monitoring programs. In the words of an environmental advisor:

> "*The biggest problem in Brazil is that they have a completely different set of laws and rules than we are used to [in Norway]. Because every single person (…) in the [Brazilian] authorities is personally responsible for his decisions. So if they make a decision and it turns out that it was not good, they might go to jail.*"

Second, according to the employees of NorthOil directly involved in the Arctic and Brazilian projects the ability to adapt the environmental monitoring machinery to very different locations is the combination of two aspects: the possibility of being out in the field and test the technology, and the collaboration with more experienced research institutions and technology vendors:

> "*It was first of all [the Brazilian project] that gave us the experience because then we were present in the field and we had four [environmental monitoring] campaigns (…) But in [the Arctic] it was the [partners who] brought in the experience and not us. So it was in collaboration that we managed to get to a concept that works better [than the first attempts]*" (ibid.)

### *4.3 Performing environmental risk*

NorthOil's experimentations soon generated space and opportunities to explore different calculations and models of risk for the selected marine resources. The results had to be meaningfully re-presented for heterogeneous but still loosely defined audiences (from the environmental experts to the drilling engineers). However, this was not a straightforward issue. The two examples below highlight the complexity of fitting risk calculation approaches into knowledge production mechanisms.

*The biomass indicator.* Echo sounders and model-inferred data can be combined to compute the concentration of biomass (e.g., fish, eggs, larvae, and zooplankton) in 3D sections of the water column every few seconds. A typical representation to visualize the measurements is the chromatogram, where data are plotted in time and coloured in different ways based on the amount of biological resources. Chromatograms can be very densely populated with data (Figure 6). They are useful to marine acoustic experts, but their granularity was deemed excessive by the environmental experts involved in NorthOil's project, who wanted to receive the results of environmental trend analysis less frequently (e.g., monthly), mostly due to the configuration of their databases. It was therefore decided to divide the water column into larger cubic sections, each associated with a *biomass indicator t*o summarize the biomass concentration inside the cubic section. The biomass indicator is obtained by collapsing some of the original sections scanned by the echo sounders into a bigger one; measures are given every hour instead of seconds. This simplified representational strategy enhanced not only the storing but also the visualisation of biomass data, moving from more than one million to less than five data entries every hour.

-- Figure 6 about here –

Simulation models are often generic tools and are based on assumptions that cannot fully account for the unpredictability of natural variation. As a result, they sometimes do not match the sensor-based measurements. This is for instance the case when simulation models are adopted to predict the potential pollution caused by well drilling activities. In this case, ad-hoc human intervention proves important to make decisions.

*The crater effect.* The effects of a planned drilling activity must always be simulated prior to the actual drilling of the new well to understand if the water current will take the drilling discharges close to sensitive resources. In a nutshell, these models are obtained by combining into a map layer the water current forecasts and the detailed drilling plan issued by the operator (see Figure 7). During the actual drilling activity, sensors are used to track some key parameters, for instance the cloudiness of the water column near the biological resources in the vicinity of the discharge point or the height and rate of particle sedimentation. It is however often the case that simulation modelling results do not match the values measured by the sensors: models might either underestimate or overestimate the spreading or deposition of the discharges. As a consequence, environmental advisors must constantly compare and contrast the modelled results and the sensor measurements to understand the reasons for the mismatches and, if possible, validate the models. Studies conducted elsewhere report that a relevant contribution is paid by the formation of cutting piles – called *crater effects*, Figure 8 – in the immediate vicinity of the drilling point (Frost et al. 2014). The craters are caused by the chemical particles in the fluids that are used to accelerate the drilling, but that can cause a faster agglomeration of the rock particles contained in the discharges. Aggregated particles therefore tend to accumulate near the borehole without propagating with the water current. In one case advisors reported how they adjusted the simulations produced by their modelling software thanks to measurements provided by one vendor (Rye and Ditlevsen 2011). Their basic idea is to incorporate the crater effect into the model through a manual workaround. After observing that the results of the simulations were one order of magnitude larger than the measured values, it seemed "a natural choice" (p. 36) for them to apply a reduction factor of 15. The new simulated results showed a "much better correspondence" with the real measurements, but, as reported, there is no theoretical justification or "rational reason behind the choice of 15" (ibid.) to simulate the amount of particles held back in the crater.

-- Figure 7 and 8 about here –

## 5. Discussion

Drawing on the general tenets of sociomateriality, our analysis engages with the specific issue of sociomaterial knowing i.e. the material circumstances of 'fact' building as demonstrated in finance (MacKenzie and Millo 2003), market formation (Pollock 2012), hotel reviewing (Orlikowski and Scott 2014) and petroleum production (Østerlie et al. 2012). Leaning heavily on these studies, our analysis highlights the following two differences that serve to characterise our notion of nested materiality.

First, our qualifier 'nested' in relation to materiality is a vehicle to forefront the actual performativity of fact building. In particular, it looks at it as it is being established, i.e. in a moment of high interaction between users and technologies. Second, nested materiality addresses the distributed, interconnected, and interacting – the infrastructural – qualities of how 'facts' are produced. The different elements (cf. Figure 9) in the subsea environmental infrastructure are recursively and mutually constitutive; they are 'nested'. According to the Merriam-Webster dictionary, "to nest" means: "to fit compactly together or within one another". In real-time environmental monitoring, materiality nests (or is made to fit) along the infrastructure (spanning natural elements; sensing devices; modelling software). These aspects of distribution, interconnection, and interaction are thus an effect of the recursive nature of infrastructures. In our empirical case, recursivity acquires a dual nature: on the one hand, the more levels of technological mediation are added, the more complexity and uncertainties are introduced (Jensen and Winthereik 2013). On the other hand, the ongoing (re-)configurations we observed are generative of new sociomaterial relations thus new opportunities to be explored (Tilson et al. 2010).

Let us now discuss the two characteristics of nested materiality in more detail.

First, we focus on technology in-the-making as opposed to black-boxed technology. This displays how users are also acting as designers and are currently giving a shape to their own infrastructure: they are interacting with it, manipulating it, testing different configurations, and discovering new relationships (e.g., between water composition and corrosivity in Brazil). We are not analysing at real-time environmental monitoring as such, but the work to establish it. Focusing on technology in-the-making makes the experimentation and co-production of facts with the technology evident. In particular, it displays how to identify facts. Clearly, not all facts are equally interesting, but who is to tell which one will be relevant? Ribes and Polk (2015) make a similar comment in his historic study of the measurements, instruments, and protocols used to trace the HIV/AIDS virus. As the causes for HIV/AIDS were not initially known, Ribes and Polk quote one informant stating that "*we were ready to handle just about any cause, as long as it wasn't aliens.*" (p. 10) In our case, the open-ended concerns for what constitutes relevant facts are highly disciplined by the constraining effect of making the most of available sensors and equipment as illustrated by the experimentation of exploiting the historic data of cod from sonars in fishing vessels (from the top) supplementing the data captured from the seabed (the bottom). This moment is one of the many in our case study that front-stage the entangled relationship between the infrastructure and human grounded expertise to conduct monitoring remotely. Our informants are aware that experience of the local ecosystem's behaviour over time is the only means to learn to interpret the new sensor readings taken upwards and compare them with those taken downwards.

Our case represents a strong emphasis on the role of the natural elements (e.g., fish, marine snow) that further compounds the challenge of identifying relevant facts. Whereas Pollock (2012) does not underline the way the market domain shapes the ranking devices in turn, in our case nature shapes both the behaviour of sensors and that of representations. Differently from Østerlie et al. (2012)'s work, the types of natural phenomena NorthOil is monitoring exist in some cases independently of the sensors (fish concentrates above the Arctic Circle to spawn anyway, whereas sand cannot jam a well without NorthOil building a well) but need to be turned into something relevant for audiences that did not consider them as a 'fact' before. The participants are constantly reflecting on new combinations of equipment to capture these facts. For instance, the intentionality of fish ("adult fish decides for itself") makes it difficult to predict their location in the water column. In contrast to fish, eggs and larvae passively drift with the ocean streams and thus need additional attention by operators: in case of an oil spill they cannot swim away. However, they are not easily detectable; their movement is inferred from software models which treat them as particles. This catches NorthOil and its advisors between having data but no models (adult fish) or models but no data (fish eggs and larvae). In this latter case nature is significant for being absent.

Second, the fact production in our case is tied to a comprehensive infrastructure, a distributed and interconnected network of sensors, communication links, and tools. Rather than being conveyed through single tools/artefacts (e.g., the market ranking device; TripAdvisor's algorithms), real-time subsea environmental monitoring spans heterogeneous layers of sensors (e.g., echo sounders, subsea cameras, light sensors), desktop systems (e.g., chromatograms, particle counters, biomass indicator, discharge simulation models), and expertise (e.g., marine biology, geology, digital electronics) – see Figure 9. Consider the example of dynamic biomass monitoring (fish, eggs, and larvae) and the calculation of the 'biomass indicator'. There exists no such phenomenon as a biomass concentration in nature. Biomass concentration is a purposeful human construct that is composed by the marine resources captured by the available sensing devices (e.g., adult cod but not cod eggs). Further, the intersection between the natural and the sensing elements (see right-hand side of Figure 9) is then fed into models (e.g., chromatogram). These need to be simplified and filtered to fit the left hand side of figure 9, i.e. the databases and software tools.

-- Figure 9 about here --

Models are recurrently used in marine environmental monitoring to produce sufficient and readable results. However, the materiality of modelling software – which is born to be as generalizable as possible – because of its generality, filters the perception of the subsea ecosystems by the experts in the control room. Human intervention in re-presenting a portion of the subsea that does not correspond to any physical or theoretical construct, as in the case of the 'crater effect', is required when models are made part of a risk assessment practice. The 'facts' are often imprecise yet sufficient for the purpose at hand. In a similar way, Edwards describes the relationship between data and (climate) models as "highly incestuous" (1999, p. 452). Representations do not depict one 'true' nature. Their purpose is rather to inscribe negotiated relationships (e.g., environmental experts and drilling engineers) and allow for knowledge to travel and be reproduced, for example to present readable results to authorities.

Our case displays a distributional variability that is not as visible in other contributions, both in terms of distance between the artefacts but also in terms of geography. Our informants spoke of "dramatically" changing the lander and reconfiguring the sensors installed in Brazil. What is done to track coral reefs in the Arctic cannot be the same as for calcareous algae in Brazil. The materiality of these creatures makes them very different. In addition, NorthOil experts had to learn about the Brazilian negotiations, quite different from those in Norway driven by political (e.g., opening the areas off North Norway) and economic (e.g., finding more subsurface resources) reasons. Therefore, the monitoring activities in heterogeneous settings are adapted and linked through the professionals' experience of the different physical conditions as well as of the socio-political ones.

## 6. Conclusions

Facts about the Arctic marine ecosystem are produced, not given. As our case of real-time subsea environmental monitoring demonstrates, *what* fact is known is invariably tied up with *how* the fact is known. Responding to the recognised under-specification of the programmatic slogan that "matter matters" (Barad 2003), our analysis makes the clear the material circumstances of 'facts'. These facts are key in deciding the future of contested, political questions around where, indeed, *if*, oil and gas operations are to operate alongside commercial fishing and environmental concerns.

The environmental impact is difficult to unpack due to the complexities of natural and technological systems (Barry 2013, p. 13). To illustrate, the government of the US in 2012 denied an international oil and gas company permission to drill subsea wells offshore North Alaska. The concluding report motivates the decision by pointing to the unique challenges associated with the Arctic area, the combination of its "environmental and weather conditions, geographical remoteness, social and cultural considerations, and the absence of a fixed infrastructure to support oil and gas activities" (DOI 2013, p. 6). Addressing the interplay between increasingly distributed and interconnected remote technologies with the environmental and social contexts is fundamental to assess the risk related to human activities. This is relevant also in terms of an early definition of responsibilities in the oil and gas industry where (in)action might lead not only to immediate but also long-term environmental damage. Our analysis is helpful in opening the black box of 'facts' to make visible their material making thus fallibility.

Table 1. Data types collected during the study, temporal extent or quantity, and themes covered.

| Data types and extent | Theme |
|---|---|
| **Participatory observations** <br> **(2-3 days a week * 2 years; field notes; hundreds of pages)** | |
| - In-office co-location with 4 NorthOil employees <br> - Internal briefing sessions <br> - Meetings with other departments <br> - 41 teleconferences (1–6 h) and workshops (1–2 days) with other NorthOil offices and the partners <br> - Informal chats | - Ongoing environmental monitoring projects <br> - Data management and work processes <br> - Stakeholders enrolment <br> - Sensor network configurations |
| **33 semi–structured interviews** <br> **(25 taped and transcribed, 10-15 pages each; 8 non-taped, 3-5 pages of notes each; avg. duration 1h)** | |
| - 23 NorthOil employees <br> - 10 employees of industrial partners | - Environmental monitoring and risk assessment <br> - Relations between ongoing and previous projects <br> - Arctic and Brazil observatories <br> - Sensor technology integration |
| **Document and corporate software analysis** <br> **(Field notes; occasional frequency)** | |
| MS SharePoint team sites (Intranet): <br>    - Internal to NorthOil <br>    - Shared with partners or vendors | - Private emails exchanged during the project <br> - Official reports and deliverables <br> - Internal notes and presentations <br> - Software requirement specifications <br> - Subsea devices technical specifications |
| Internal information systems | - Corporate Geographical Information System <br> - Test version of environmental risk modelling software <br> - Public web portal for real-time environmental data <br> - Work processes repositories |

Table 2. Our interpretive template with three moments of performativity observed during our analysis. For each moment, we provide an empirical illustration drawn from Section 4 and an excerpt from the interviews.

| Moment of performativity | Empirical illustration | Excerpts |
|---|---|---|
| **Nature** | The way a coral structure is monitored depends on its health and the fish species it attracts, how it fits the available subsea camera lens, on the even terrain to position the sensor structure so that it would not flip over. | *"[Y]ou are basically looking for something as interesting as possible to put in one photographic frame (…) it would be good if the coral reef could also attract fish species (…). [W]e had the necessity to have a flat position [to prevent the sensor support structure to fall over]".* |
| **The operational site** | Environmental monitoring practices are part of local socio-political practices, where the physical materiality of the subsea field gets entangled with interests and normative frameworks. | *"[Environmental monitoring] needs to be based on the sort of political regime, basically the regulations in the country where you are operating."*<br><br>*"[T]here was a deal of problems [in Brazil] with corrosion and short circuits because the environment in [Brazil] is extremely corrosive. We had not taken that into account."* |
| **Environmental risk** | High-level representations of online environmental datasets are generated and manipulated to produce simplified representations of the risk for the targeted resources. | *"These pictures indicate that a large part of the discharge (…) may be "trapped" inside the "crater" of the pile formation (…) This simulation attempted to take into account the "crater effect" (…) The amount of discharge was reduced with a factor of 15 (…) There is no rational reason behind the choice of a factor 15 of the particle matter to be held back in the crater [, it] seems a natural choice."* (Rye and Ditlevsen 2011) |

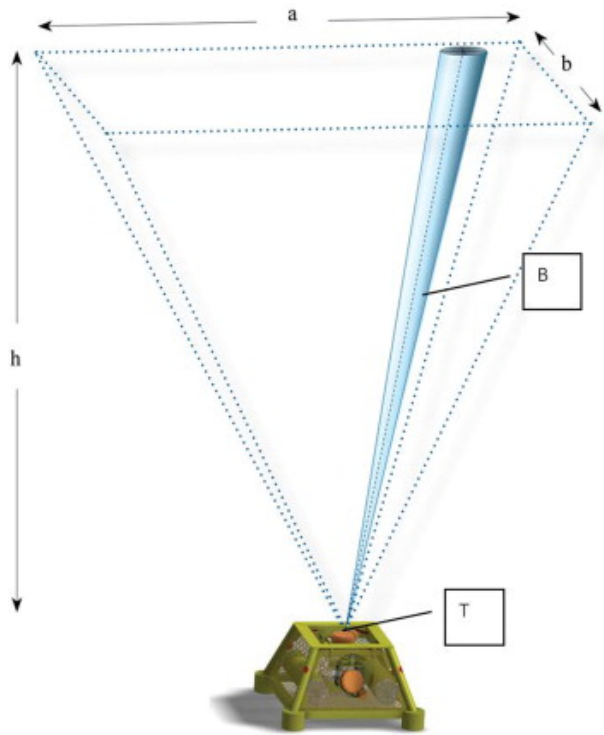

Figure 1. Example of a subsea lander (bottom) and the spanning area of the echo sounders (Source: Godø et al. 2013)

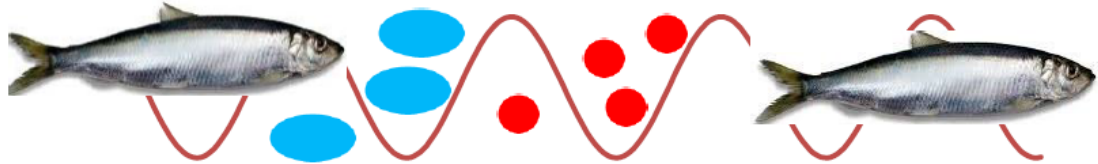

Figure 2. Exaggerated schematic representation of sound wave emitted by an acoustic sensor with reference to the size of adult fish, eggs (blue), and larvae (red).

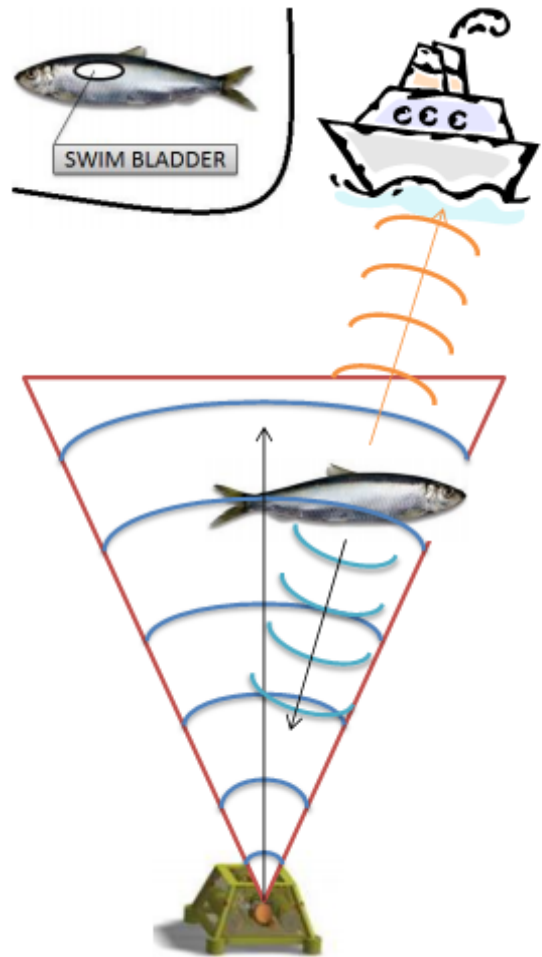

Figure 3. Top, left: position of the swim bladder. Bottom and right: exaggerated schematic representation of the orientation of the echo sounder in Arctic observatory (purple circle) compared to the traditional use from boats.

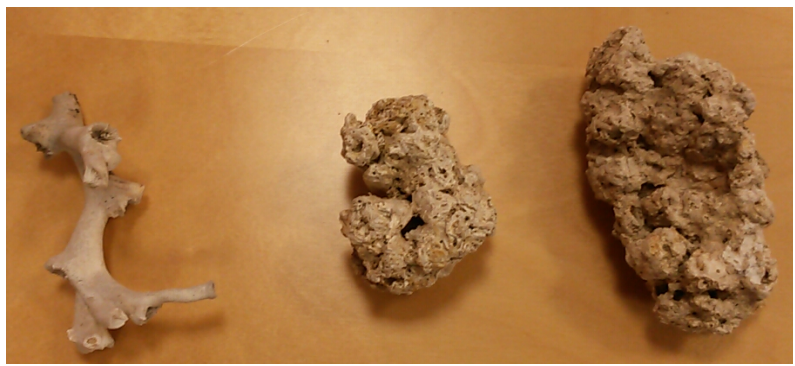

Figure 4. Left, a portion of dead coral; center and right: two portions of dead agglomerates of dead calcareous algae.

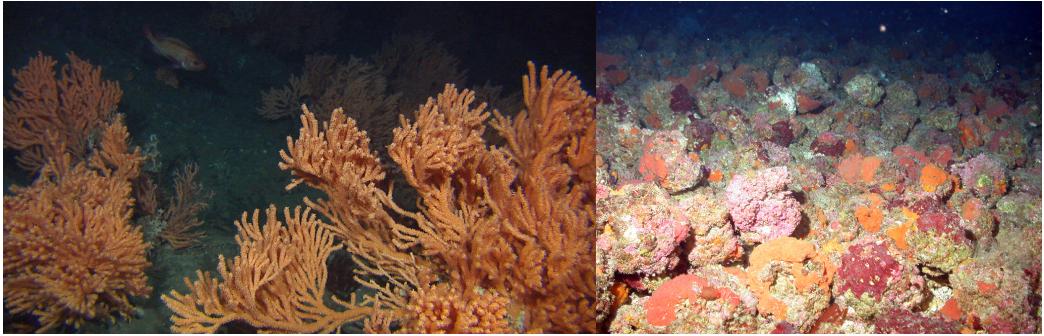

Figure 5. Left: a coral forest (source: http://ww.mareano.no). Right: nodules of calcareous algae (source: http://flowergarden.noaa.gov).

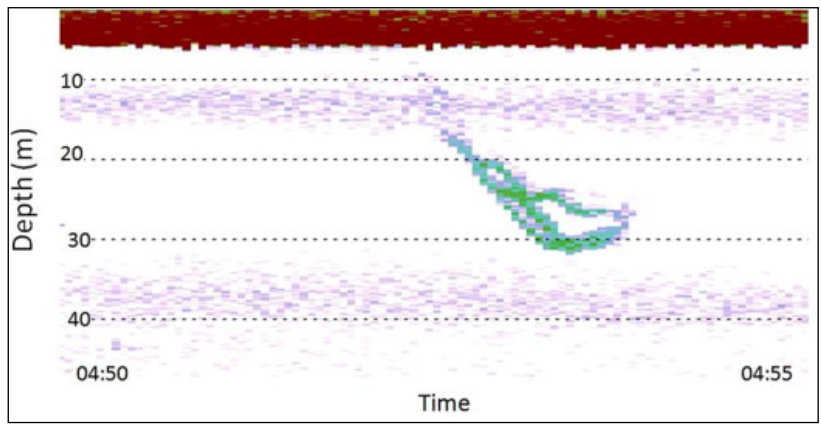

Figure 6. Example of chromatogram (Source: Godø et al. 2013).

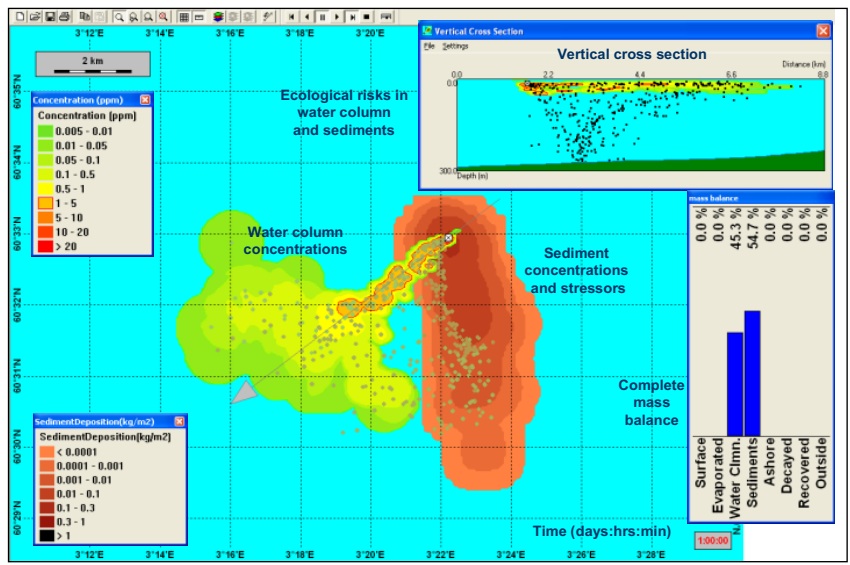

Figure 7. Map of the predicted plume of particles generated during the drilling of a well with reference to the water currents (Source: Rye and Ditlevsen 2011).

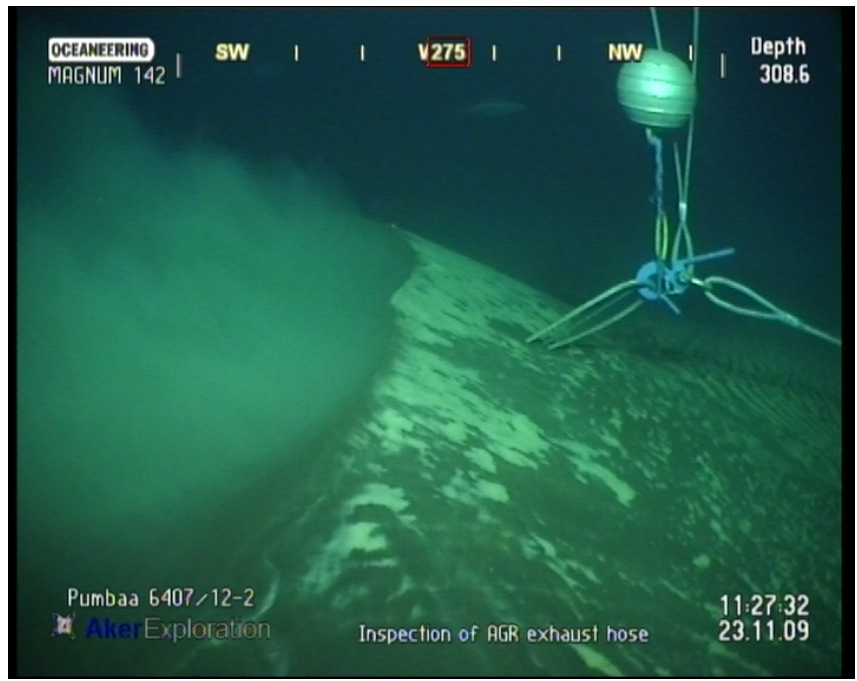

Figure 8. The formation of a crated during drilling (Source: Rye and Ditlevsen 2011).

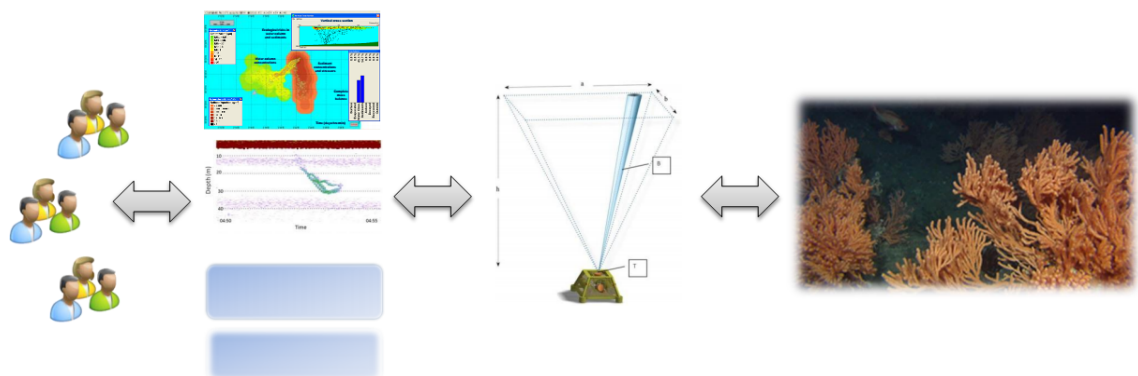

Figure 9. A schematic representation of NorthOil's infrastructure unfolding along a distributed network of digital devices spanning from the seafloor to the users in the control center.